\documentclass{elsart5p}
\usepackage{graphicx, amsmath, amssymb, natbib}
\begin{document}
\begin{frontmatter}
%
\title{Package models and the information crisis of prebiotic evolution
}
%
\author[IFSC]{Daniel  A. M. M. Silvestre,}
\author[IFSC]{Jos\'e F. Fontanari\corauthref{CA}}
\corauth[CA]{Corresponding author: fontanari@ifsc.usp.br}
\address[IFSC]{Instituto de F\'{\i}sica de S\~ao Carlos, Universidade de S\~ao Paulo, Caixa Postal 369, 13560-970 S\~ao Carlos SP, Brazil}
%

%
\begin{abstract} 
 
The coexistence between different types of templates has been the choice solution to the information crisis
of prebiotic evolution, triggered by the finding that a single RNA-like template cannot 
carry enough information to code for any useful replicase. In principle, confining $d$ distinct templates
of length $L$ in a package or protocell, whose survival depends on the coexistence of the
templates it holds in,  could resolve this crisis provided that $d$ is made sufficiently large. Here we
review the  prototypical package model of \citet{Niesert1981} which guarantees the greatest possible 
region of viability  of the protocell population, and show that this model, and hence the entire 
package approach, does not
resolve the information crisis. This is so because to secure  survival the total information
content of the protocell, $L d$,  must tend to a constant value that depends only on the
spontaneous error rate per nucleotide of the template  replication mechanism. As a result, an increase of
$d$ must be  followed by a decrease of $L$ to ensure the protocell viability, so that 
the net information gain is null.
\end{abstract}
%
\end{frontmatter}
\section{Introduction} \label{intro}

The information crisis in prebiotic or chemical evolution, as envisaged by \citet{Eigen1971} in the 1970s, 
is grounded in two key observations: 
(i) the length of a replicating polymer (i.e., RNA-like template) is limited by the fidelity of replication 
\citep{Eigen1971}, and (ii) templates that differ from each other such that their differences reflect on their replication rates
cannot coexist in a purely competitive setup \citep{Swetina1982}. Despite some criticisms about the biological relevance of 
the information crisis 
(see, e.g., \citet{Wiehe1997, Wilke2005}), the seminal work of Eigen has raised true questions about the onset of life,
fueling the  still unresolved debate on the  mechanisms that explain how the primeval replicators coped
with the loss of information resulting from  competitive exclusion and recurrent mutations.

There are two competing  scenarios that seek to solve the prebiotic information crisis by achieving template coexistence. 
In principle, template
coexistence solves the information crisis because the total information content of the template pool
is the product of the number of different templates and the maximum information coded per template, 
provided the template types have the same concentration.
The first scenario is the hypercycle, a closed reaction scheme in which each replicating polymer aids 
in the replication of the next one, resembling a current motif in biochemical metabolic pathways \citep{Eigen1978}. 
This scheme requires that the primordial replicators act both as templates and as cross-catalytic replicases, but 
this key assumption encounters strong criticism. In fact, since natural selection is responsive to replication rates only, 
one can expect  it to make each element of the hypercycle  a better target for the replicase 
(or a better replicase to itself), but one cannot expect  natural selection to drive 
the cross-catalytic activity of the hypercycle components in the same way. 
Therefore, the likely fate of each replicase is to mutate into the  so-called parasites 
-- molecules similar to the modern genomic polymers which lack catalytic activity \citep{Smith1979, Bresch1980}. 
In addition,  hypercycles with a large number of different templates are subject to large fluctuations 
in the concentrations of its members, rendering it  susceptible to extinction by stochastic effects \citep{Nuno1994, Campos2000}. 
Nonetheless, the hypercycle model has received substantial empirical support regarding its basic premises as 
catalytic activity in a variety of biological molecules, including DNA and RNA, and the experimental synthesis of 
minimal self-replicating systems are now well established 
(see \citet{Paul2004}).

The alternative prebiotic scenario to address the template coexistence issue is the 
so-called package model approach, in which the templates are confined in packages termed
protocells  or prebiotic vesicles
\citep{Gabriel1960, Bresch1980, Niesert1981, Niesert1987, Silvestre2005, Fontanari2006}. 
The basic assumption here is that a nonspecific replicase is codified and assembled
by a number $d$ of distinct functional templates. Template replication is possible only with the aid of
the replicase which, in turn, can be put together only when all template types 
are present in the package.

The combined dynamics of templates and protocells is in many aspects similar to the metapopulation dynamics 
used in the study of patches in ecology (see, e.g., \citet{Keymer2000}). The distinguishing features of the package models
are the random drift in the template dynamics due to the finite capacity of the protocells,
and the assortment load resulting from the stochastic fission of the protocells. These very features that
make the dynamics interesting, also make it much less amenable to analytic treatment.
In fact,  as compared with the hypercycle and quasispecies models for which the deterministic chemical
kinetics equations suffice to describe most of the relevant phenomena
(see, however, \citet{Alves1998, Campos1999}), package models like the classical model of \citet{Niesert1981}
or the stochastic corrector model  of \citet{Szathmary1987} are not so well characterized and only recently 
slightly simplified versions of the original models 
were analyzed thoroughly \citep{Grey1995, Zintzaras2002, Silvestre2005, Fontanari2006, Silvestre2007}. 

Our main reservation about the package models premises concerns its group selectionist flavor. One way or the other, 
all aforementioned package models assume  that the growth rate 
of the protocell lineage  depends on the  composition of the template
population confined within each protocell. Although at first sight 
this may seem a reasonable assumption, it is not supported by experimental evidence:  recent experiments
with model protocells show that the growth of lipid vesicles does not depend on the nature of the 
substances they hold in \citep{Hanczyc2003, Chen2004, Chen2006, Zaher2007}. A more plausible assumption
is to admit that protocell fission is triggered when the  total concentration of the  templates
reaches some critical value so that the  integrity of the membrane is compromised \citep{Chen2006}. 

Interestingly, the package model proposed in the pioneer work of \citet{Niesert1981} conforms to this more realistic scenario, as 
no  connection is made between template composition and the mechanism of protocell fission. Alas, for purely technical reasons
-- to keep the protocell lineage to a size that could be manageable by  their computational resources -- those authors have introduced
an extraneous ingredient into their model, namely,  a prospective value which essentially gauges the survival
probability of a protocell according to its template composition. This prospective value plays exactly the same
role as the group selection pressure in models where a direct relation between template composition and
protocell fission (or protocell reproduction) is made explicit \citep{Zintzaras2002, Silvestre2005, Fontanari2006}.
To handle and quantify the unlimited growth of the
protocell population we borrow 
a powerful tool from statistical physics, namely, the spreading or epidemic analysis 
used to characterize  nonequilibrium phase transitions in lattice models \citep{Grassberger1979}. 
This technique suits
particularly well to the task of locating the boundaries in the  space of parameters
that separate the subcritical regime, where the extinction of the lineage is certain,
from the supercritical regime, where the probability of survival of the lineage  is nonzero
(see, e.g., \citet{Rosas2003}).

Since ingredients such as prospective values for protocell survival or  protocell fitness 
commonly used in the definition of  package models are  not supported by  experimental evidence,
we find  the reexamination of the original version of the model of 
\citet{Niesert1981} absolutely essential. More importantly, because viable protocells (i.e.,
protocells that contain the $d$ functional template types) are
allowed to reproduce unrestrainedly, the region of viability of the metapopulation is the greatest possible
in this model. Any other package model that assumes protocell competition by assigning fitness values to the
protocells according to their template compositions will exhibit  a smaller region of viability. 

We have found that the confinement of templates in packages does not solve the information crisis of prebiotic evolution because, 
to guarantee the viability of the protocells, the  total information content of each  package must remain constant, i.e., the
product between the template lengths $L$ and the number of distinct templates $d$ is a constant that  depends, essentially, 
only on the spontaneous error rate per nucleotide. Hence there is no information gain of increasing the number of coexisting 
templates types since their lengths must decrease proportionally. 

The rest of the paper is organized as follows.  In Sect.\ 
\ref{model} we review the package model of \citet{Niesert1981} and  in  Sect.\ \ref{Spreading} we briefly
describe the spreading analysis technique. The results are presented in a crescendo of difficulty  from Sects.\
\ref{sec:diversity} to \ref{sec:lethals}. In particular, Sect.\ \ref{sec:diversity} is
dedicated to the study of the limit of perfect replication fidelity; Sect.\ \ref{sec:parasite} considers
the effect of mutations to the so-called parasite class, which are essentially nonfunctional templates; and
Sect.\ \ref{sec:lethals} the effect of mutations to the lethal class, which
are precipitating agents that prevent the assemblage of a  replicase. 
Finally, Sect.\ \ref{sec:conclusion} is devoted to our concluding remarks.

\section{The package model of \citet{Niesert1981}}\label{model}

According to the model of \citet{Niesert1981} we 
consider a metapopulation composed of a variable number of packages or protocells, each of  which encloses a certain number of templates
(RNA-like molecules) and a few polypeptides. 
Since all our simulations  begin with a single mutant-free protocell we use the words metapopulation and protocell
lineage interchangeably  throughout   this paper.
There are  $d$ distinct types of functional templates which, when present in the same protocell 
are capable of assembling a nonspecific replicase of finite processivity 
and finite fidelity of replication. As pointed out by \citet{Niesert1981}, this kind of 
primitive translation is achieved by cooperation of RNAs with t-RNA character and an RNA with
messenger function. Here we define the processivity $\Lambda$ of the replicase as  the number of template copies it
produces in some convenient unit of time,  so $\Lambda$ can be viewed as a measure of replication efficiency as well. 
This definition can be made equivalent to the usual definition of processivity as 
the average number of nucleotides added by a  replicase per 
association/disassociation with the template if  $\Lambda$ is allowed
to take on noninteger values and all templates have the same length.
We recall that the  replicase can only be formed in the presence of all $d$ types of template. If 
a protocell lacks a single  functional template type it is considered unviable and 
discarded from the protocell population. In addition, all template types display identical targets to
the replicase, which results in a neutral replication process, i.e., all template types have the
same replication rate.

Since any plausible replication process is susceptible to errors, we must take into account the possibility
of  mutations. 
Two types of mutations are allowed in the model. First, there are mutations that produce non-functional templates 
which contribute nothing to the assembling of the replicase but keep their replication capability unharmed. These
mutants are termed parasites. Second, there are lethal mutations which generate molecules that prevent the assembling
of a replicase or, equivalently, produce a non-functional replicase. In any case, an offspring
protocell that contains a mutant of this kind is rendered unviable. 
Mutations to the parasite type occur with probability $u$ whereas mutations to the lethal type occur with probability $v$. 
Reverse mutations or mutations between functional template types are not considered. The life cycle of the protocells consists 
of three events -- template replication, 
protocell fission and protocell demise -- which take place  in this order and are described in the following.

\subsection{Template replication}

This process, which describes the replication of the templates inside the protocells, 
is regulated by one of the crucial parameters of the model, namely, the number of replicated molecules between two 
vesicle fissions ($\Lambda$). As already pointed out, this quantity is the processivity of the replicase, i.e., 
the number of template copies the replicase can produce in a unit of time, taken here as the time 
between two consecutive fissions. The replication process is implemented as follows. 
For each protocell we choose a template at random and replicate it, 
returning both the original template and its copy to the protocell, and  this procedure is repeated until 
the processivity of the replicase  is reached, i.e., it is repeated exactly $\Lambda$ times. If the
randomly selected template is a functional template then the copy will become a parasite with probability $u$
or a lethal with probability $v$. Hence the probability that the copy is perfect is $1 -u - v$.
If the selected template is a parasite, then the copy will become a lethal with probability $v$. Finally,
in the case a lethal is selected, then the copy will also be a lethal. As a result of this procedure, exactly
$\Lambda$  new templates are added to the protocell.
We note that in the absence of mutations this process  is exactly the Polya's urn scheme \citep[Ch. V.2]{Feller1968}.
The template replication procedure is repeated for all protocells in the metapopulation. In what follows we will refer to the
number of templates inside a protocell as the size of the protocell.

\subsection{Protocell fission}

After template replication, the protocell is in  excess of molecules since we assume that the membrane is 
impermeable to templates. As  template replication
is much faster than protocell fission, we  envisage a scenario where
fission begins at about the same time as the template replication cycle but terminates only after 
the replicase has exhausted its copying capacity. This means that the overall osmotic pressure tolerated 
by the protocell membrane is  of the order of $\Lambda$, so that   the fission process is triggered
whenever the protocell size exceeds $\Lambda$. For example, if the size of
a just generated daughter protocell is $n > \Lambda$ then the fission process starts but, by the time
it is concluded, the protocell has already reached the size $ n + \Lambda$.

Experimental models 
of lipid vesicles exhibit  a variety of mechanisms for the fission process \citep{Hanczyc2004}. Here we choose a 
mechanism that splits the mother protocell in two daughters and  distributes
templates of the mother vesicle  between the two offspring at random.  
The random assortment of templates to the daughter vesicles  may cause the loss of essential genes 
for survivorship (i.e., the assortment load), thus producing unviable protocells. 
To be more precise, if $S$ is the size of the mother 
protocell just before fission then the probability that the daughters have size 
$s$ and $S-s$  is given by the binomial  $\binom{S}{s}2^{-S}$.  

\subsection{Protocell demise}

The viability of a daughter protocell is guaranteed provided that (i) it carries no lethal mutants,
and (ii) it has at least one copy of each type of functional  template. Any protocell lacking one of those templates 
or carrying a lethal mutant is dismissed because it is incapable of producing a working replicase. 

\subsection{Metapopulation dynamics}

The sequence of the three processes stated above defines the time unit of the  metapopulation dynamics. 
>From these processes it is clear that the size of  the metapopulation  (i.e. the number of protocells) can, in some cases, 
increase without bounds. In fact, the metapopulation dynamics can be viewed as a branching  process with denumerably 
many types \citep{Kimmel2002}.
As pointed out before, to circumvent this `technical' difficulty which rapidly saturates the computer resources of the time, 
\citet{Niesert1981} have opted to discard supernumerary protocells according to an arbitrary 
prospective value which essentially gauges the odds of a protocell to leave viable descendents (see Sect.\ \ref{sec:parasite}).
In practice, using such a prospective value to eliminate the least promising protocells introduces an additional criterion for 
protocell demise, which can be interpreted as a selection pressure acting  at the protocell level which favors protocells
with high prospective values.  Since the survival probability of the protocell begins to depend on specific details 
of their template compositions, the resulting model becomes very similar to the group selection  models
proposed to explain template coexistence (see, e.g., \citet{Fontanari2006}).
Interestingly, \citet{Niesert1981}  do not mention group selection or multi-level selection: although they state that  
the  unit of selection is the entire package, they overlooked the conflict between package and  template interests, 
especially the parasites.   Here we do not use 
any prospective value to control the metapopulation size and the only criteria for elimination of a protocell 
is the presence of a lethal mutant or  the lack of any type of functional template.

Although the number of protocells may grow unboundedly, the number of templates inside each protocell
(i.e., the protocell size) fluctuates around a finite value, though, strictly, a viable protocell can
assume any size value 
in the range $d,\ldots,\infty$. In fact, starting from whatever conditions the 
\textit{average} number 
of templates per package will approach the processivity of the replicase very quickly. 
Consider, for example, the average number of templates $n_{t}$ in a protocell at time $t$. It is related to the average number of 
templates of its mother protocell $n_{t-1}$ by the  recursion equation
\begin{equation}\label{eq1a}
n_{t}=\frac{(n_{t-1}+\Lambda)}{2}
\end{equation}
whose solution is simply
\begin{equation}\label{eq1b}
n_{t}=\frac{n_0 -\Lambda}{2^{t}}+\Lambda ,
\end{equation}
where $n_0$  is the size of the vesicle that originated the lineage. Hence, regardless of
the ancestor size,  after a few  generations the average size of the protocells will equal the
processivity value of the replicase. 

The ultimate goal of the analysis of the metapopulation dynamics is to determine the regions in the
space of the parameters $d$ (number of functional templates), $\Lambda$ (processivity of
the replicase),  $u$ (probability of mutation to parasites) and $v$ (probability of mutation to the
lethal type)
where the protocell lineage has a nonzero probability of thriving. To achieve that we resort solely to 
brute force simulations of the metapopulation dynamics as will be described in the forthcoming sections.
A slightly modified version of this model, in which  template replication takes place simultaneously through
a Wright-Fisher process and the daughter protocells have fixed size $\Lambda$, yields to an analytical approach
\citep{Silvestre2007}. In this contribution, however, we have opted to stick to the original model of 
\citet{Niesert1981} so as to complement and complete that classical work on the theory of prebiotic evolution.

\section{The spreading analysis}\label{Spreading}

The spreading or epidemic analysis put forward by \citet{Grassberger1979} is the
preferred  technique of statistical physics to locate and characterize equilibrium as well as
non-equilibrium phase transition lines. The method is tailored  to population dynamics problems
that exhibit an absorbing state (subcritical regime), where the protocell population has gone
extinct, and an active state (supercritical regime) where the population undergoes exponential growth. 
Separating these two regimes there is the critical regime where the population growth (if any) is
sub-exponential, usually a power law in the time variable.

The basic idea of the spreading analysis is to follow the evolution of the lineage that sprouts from a single mutant-free protocell of size 
$\Lambda$, in which the functional templates are evenly represented. For each time $t$  we carry out from  $10^6$ to $10^9$
independent runs, all starting with the same ancestor protocell. The number of runs depends on the values of the control parameters,
and is chosen so as  to guarantee a representative number of surviving samples at any given time. Here we focus on the 
time dependence of two key quantities: the average number of protocells $ N_t$  and the survival probability of the lineage $P_t$
calculated at time $t$. 
Clearly, $P_t$ is simply the fraction of runs for which there is at least one protocell in the lineage at time $t$. We note that in 
the calculation of $N_t$  we take an average over all runs, including those for wich the lineage has already gone extinct at time $t$.

Let $\Delta$ be a real variable that measures the distance to the critical point. For example, if the transition from the
absorbing to the active phase takes place at the mutation probability $u = u_c$ when all other parameters are
held fixed then $\Delta = u_c - u$.   Close to this transition point, i.e., for $\Delta \approx 0$,  and
for sufficiently large $t$,
we expect that the average number of protocells and the survival probability obey the scaling hypothesis
\begin{eqnarray}
N_t & \sim &  t^\eta \phi_N \left ( \Delta^\nu t \right ) \label{scaleN} \\
P_t & \sim &  t^{-\delta} \phi_P \left ( \Delta^\nu t \right ) \label{scaleP}
\end{eqnarray}
where $\eta$, $\delta$ and $\nu$ are non-negative critical dynamic exponents, and $\phi_N$ and $\phi_P$ are
universal scaling functions, usually exponential functions \citep{Grassberger1979}. At the critical
point $\Delta = 0$, a log-log plot of $N_t$ (or $P_t$) as function of $t$ yields a straight line, the
slope of which yields the dynamic exponent $\eta$ (or $\delta$). Upward and downward deviations   from
the critical straight line indicate supercritical and subcritical behaviors, respectively. 
The mere observation of these deviations allows a very precise estimate of the value the 
critical parameter $u_c$. 

A particularly important quantity, as far as the characterization of 
branching processes is concerned, is the  ultimate probability of survival of the lineage, defined as
\begin{equation}\label{P_eq}
\Pi = \lim_{t \to \infty} P_t
\end{equation}
so that $\Pi = 0$ in the critical and subcritical regimes ($u \geq u_c$), whereas $\Pi > 0$ in the
supercritical regime ($u < u_c$). In the latter case, for $u$  sufficiently close to $ u_c$
we have another scaling relation, $\Pi \sim \Delta^\beta$ where $\beta$ is a static critical exponent.
Since Eq.\ (\ref{scaleP}) holds for $t \to \infty$ in the sub as well as in the supercritical regime
the critical exponents $\delta$, $\nu$ and $\beta$  must not be all independent. To see that we 
rewrite Eq.\ (\ref{scaleP}) in the supercritical regime as $P_t \sim \Delta^{\nu \delta} \psi_P \left (\Delta^\nu t \right ) $ where
$\psi_P (x) =x^{-\delta} \phi_P (x)$ is an arbitrary function that tends to a finite positive value when $x \to \infty$.
It follows then that $\beta = \delta \nu$. The precise estimate of these exponents is a major concern of
a branch of the statistical physical that deals with  branching-like processes in lattice models (see, e.g., \citet{Marro1999}).
The branching process  we consider here can be viewed as a mean-field-like version of similar processes in lattice models and so
the critical exponents are the mean-field ones: $\eta =0$, $\delta = -1$, and $\nu = \beta = 1$.
The \textit{a priori} knowledge of these values can be useful to validate our estimate of the threshold locations,
which are actually our main concern in this paper.

\section{The cost of diversity}\label{sec:diversity}

The information crisis of prebiotic evolution revealed by the quasispecies model is essentially a
consequence of the limited fidelity of replication of the templates with the resulting steady accumulation of
mutants in the population (the  mutational load in the population genetics jargon). In  package models
there is an additional threshold phenomenon which, even in the absence of
mutations, limits the amount of information stored in the molecular pool: the limited processivity of
the replicase  bounds the number of distinct template types that can coexist in a protocell. We refer to
this new information crisis as the cost of diversity. Deterministic models, such as the  
quasispecies and the hypercycle models, assume that the processivity of
the replicase is essentially infinite but a recent estimate of this parameter in artificially selected 
RNA polymerase ribozymes yields   a processivity on the order of a few tens of nucleotides \citep{Zaher2007}, 
which is insufficient to produce even the  shortest functional ribozyme known, the hairpin ribozyme \citep{Kun2005}. 
To better characterize the processivity threshold 
in this section we focus in the case in which the replication fidelity is perfect, i.e., $u=v=0$.

Although some features of the primordial replicase such as its specificity and  processivity are  key elements 
in most prebiotic scenarios, their roles are rarely punctuated in the literature.
Notable exceptions are \citet{Michod1983} who
has argued that the lack of specificity of the primordial replicase gives the templates an altruistic-like character
which could be maintained by the confinement of the templates in packages, and
\citet{Niesert1981} who have, in turn, shown that a high processivity is necessary
to  make up for the assortment load and so to guarantee the coexistence of a few functional templates
in the protocell.

\begin{figure}
\centerline{
\resizebox{0.75\columnwidth}{!}{\includegraphics{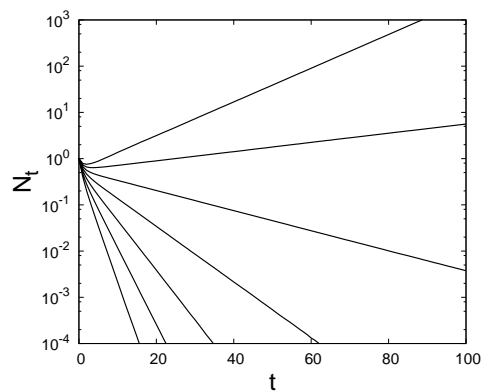}}}
\caption{Average number of protocells in a lineage originated in a single  well-balanced and
mutation-free founder. The parameters are $d=4$ and (from bottom to top) 
$\Lambda=4, \ldots, 10$. The data shown for each $t$ represents the average over $10^7$ independent runs.
The semi-logarithmic scale facilitates the observation of the exponential increase and decrease
of the lineage size in the sub and supercritical regimes.}
\label{fig:1}
\end{figure}

\begin{figure}
\centerline{
\resizebox{0.75\columnwidth}{!}{\includegraphics{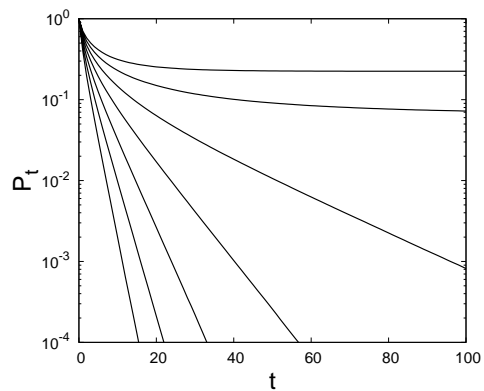}}}
\caption{Same as Fig. \ref{fig:1} but for the survival probability. For large $t$ this quantity
tends to a constant value, the ultimate survival probability $\Pi$. The minimum value of the processivity
$\Lambda$  that guarantees $\Pi > 0$ is $\Lambda_c = 9$ in this case.
}
\label{fig:2}
\end{figure}

To obtain the minimum value of the processivity $\Lambda$ of a replicase that needs $d$ functional
templates for its assembling we use the spreading technique as illustrated in Figs.\ \ref{fig:1}
and \ref{fig:2} for the case $d=4$. Given the discrete nature of $\Lambda$, the critical regime
is absent in this case and so one observes a discontinuous jump from the typical subcritical behavior
(both $N_t$ and $P_t$ vanish exponentially with increasing $t$) to the supercritical regime
($N_t$ increases exponentially with increasing $t$ whereas $P_t$ approaches the ultimate survival
probability $\Pi$). We recall that the calculation of $N_t$ includes the extinct runs as well
and this is the reason that $N_t$ can take on values  smaller than  $1$ in the subcritical regime
(note that $N_0 =1$).


\begin{figure}
\centerline{
\resizebox{0.75\columnwidth}{!}{\includegraphics{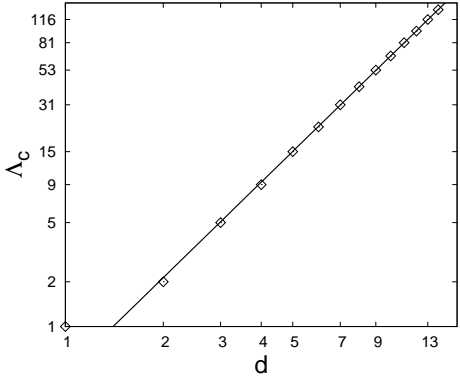}}}
\caption{Logarithmic plot of  the minimum value of the processivity $\Lambda_c$ above which the protocell lineage has a finite
probability of escaping extinction against the diversity $d$. The extinction  of the lineage is certain for $\Lambda < \Lambda_c$. 
The solid  line is the fitting $\Lambda_c = d^2/2$. The parameters are $u=v=0$.}
\label{fig:3}
\end{figure}

Repetition of the spreading analysis for different values of the diversity $d$ allows us to obtain the
dependence of the minimum processivity $\Lambda_c$ with $d$. The results are summarized in Fig.\ \ref{fig:3}, which
essentially shows that  for $d$ not too small the data is described very well by the fitting
$\Lambda_c = d^2/2$. For $\Lambda < \Lambda_c$, the replicase is unable to produce sufficient copies of the templates to compensate for the 
stochastic loss of functional template due to the assortment load, and so the lineage goes extinct with probability one.
For $\Lambda \geq \Lambda_c$, the lineage has a nonzero probability of ultimate survival.

As pointed out by \citet{Silvestre2007}, there is a simple combinatorial argument to
explain the scaling $\Lambda_c \sim d^2$ at the critical boundary in the error-free replication
case, that goes as follows. Assuming that the typical size of the protocells is $\Lambda$ then  the number of different types of
viable protocells  is 
\begin{equation}
\left ( \begin{array}{c} \Lambda - 1\\ d - 1 \end{array} \right ),
\end{equation}
whereas the total number of protocell types is 
\begin{equation}
\left ( \begin{array}{c} \Lambda + d - 1\\ \Lambda \end{array} \right ). 
\end{equation}
The logarithm of the ratio $r$ between these two quantities can be written as  
\begin{equation}\label{r}
\ln r =  \sum_{i=1}^{d-1} \ln \frac{ 1 - i/\Lambda}{1 + i/\Lambda } \approx -\frac{2}{\Lambda} \sum_{i=1}^{d-1} i
\approx -\frac{d^2}{\Lambda},
\end{equation}
from where we can see that the only way to obtain  nontrivial   values of  this ratio (i.e., $r \neq 0,1$) for large $\Lambda$ and $d$
is to assume that  $d^2/\Lambda$ remains of order of 1, which yields $r \sim \exp \left ( - d^2/\Lambda \right )$. Clearly, 
$r \to 0$  corresponds the subcritical regime since  the fraction of viable protocell types is vanishingly small in this case, whereas
$r \to 1$ corresponds to the supercritical regime as practically all protocell types are viable.

The relation $\Lambda_c \sim  d^2/2$ is the same as  
that found in the original work of \citet{Niesert1981} as well as in the variant of \citet{Silvestre2007}, and it poses another 
challenge to the primordial protocells, for  which no mechanism of segregation can
be assumed to exist:  the need to overcome the efficiency lower bound $\Lambda_c$ in order to
maintain a given number of distinct functional templates. This threat to prebiotic evolution was aptly termed Charybdis  
by \citet{Niesert1981} in reference to the monster of the Greek mythology.

\begin{figure}
\centerline{
\resizebox{0.75\columnwidth}{!}{\includegraphics{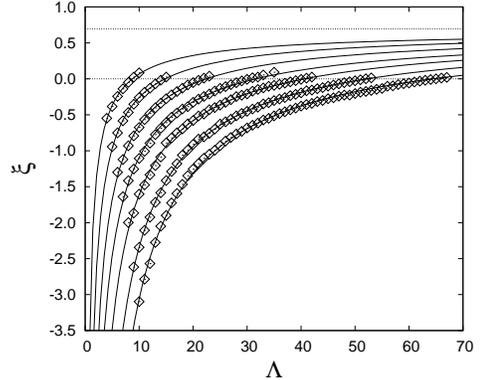}}}
\caption{ Asymptotic growth rate of the lineage  as function of the replicase processivity in the error-free case for (from top to bottom) 
$d=4,\dots,10$. The solid lines are the fittings $\xi = \ln 2 + a \Lambda^\theta$ with $\theta < 0$. The value of  $\Lambda$
above which  $\xi$ becomes positive defines the limits to the realm of viability of the lineage.  The horizontal lines
are $\xi = 0$ and $\xi = \ln 2$.
}
\label{fig:4}
\end{figure}

There is yet another difficulty that passed unnoticed in previous works and that makes
the  diversity cost transparent. From Fig.\ \ref{fig:1} we can easily see that $N_t \sim \exp \left ( \xi t \right )$
for large $t$  where $\xi = \xi \left ( \Lambda,d \right )$ is  the asymptotic growth rate of the metapopulation,
shown in Fig.\ \ref{fig:4}. The point here is that for $\Lambda$ fixed, the growth rate decreases
with increasing diversity, which  means that there will be a selective pressure to increase $\Lambda$ while $d$ is
kept to some minimum possible value. A way out of this conundrum would be to assume that  the processivity 
of the replicase depends somehow on the information content of the template pool (i.e., $\Lambda = \Lambda \left ( d \right) $) 
so that it would be impossible for a system of $d$ template types to assemble a replicase with  
processivity $\Lambda \gg \Lambda_c \sim  d^2/2$.

\section{The parasite load}\label{sec:parasite}

To better realize the implications of an imperfect replication mechanism, we will  focus first
on the mutations to the parasite type only, i.e., $u > 0$ but $v =0$.
The accumulation of parasites in a  protocell does not make it unviable, but in the long run  it can reduce the efficiency of
the replicase, leading to the underproduction of functional template copies that could  counterbalance
the disruptive effect of the assortment load. (We recall that the parasites have the same replication rate as the
functional templates). 

\begin{figure}
\centerline{
\resizebox{0.75\columnwidth}{!}{\includegraphics{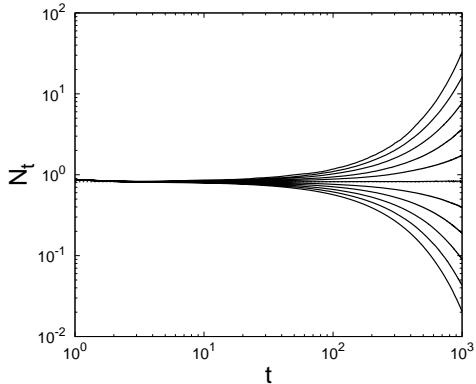}}}
\caption{Logarithmic plot of the average number of protocells in a lineage 
 for several values of the parasite mutation probability 
(top to bottom) $u= 0.180,0.181, \ldots,0.190$. The critical mutation probability
is $u_c \approx 0.185$ and corresponds to the horizontal straight line ($\eta =0$).
The parameters are $d=2$, $\Lambda=2$, and $v=0$. }
\label{fig:5}
\end{figure}

\begin{figure}
\centerline{
\resizebox{0.75\columnwidth}{!}{\includegraphics{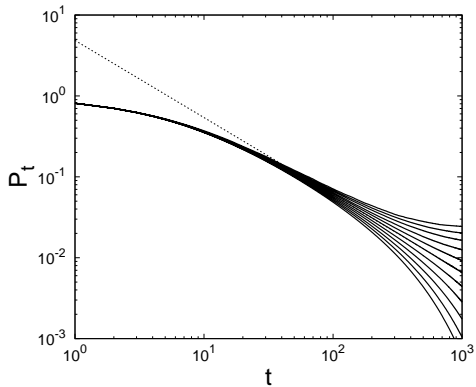}}}
\caption{Same as Fig.\ \ref{fig:5} but for the survival probability. The dashed straight
line is the fitting $P_t \sim t^{-1}$ of the critical curve.}
\label{fig:6}
\end{figure}

\begin{figure}
\centerline{
\resizebox{0.75\columnwidth}{!}{\includegraphics{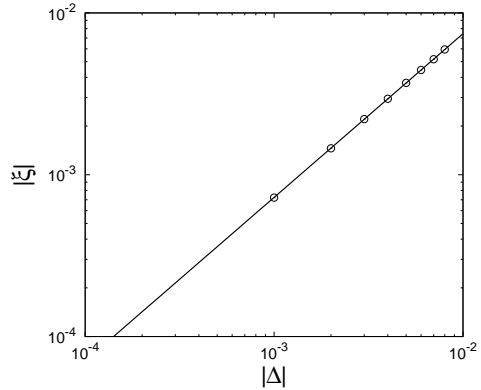}}}
\caption{Logarithmic plot of the absolute value of the time decay constant $\xi$  against $\mid \Delta \mid$ for the data
shown  Fig.\ \ref{fig:6}. The  slope of the straight line used to fit the data is $\nu = 1.01 \pm .01$.
}
\label{fig:7}
\end{figure}

Since the mutation probability $u$  varies continuously in the range $\left [ 0,1 \right ]$, in this section
we can finally appreciate the effectiveness of the spreading analysis to locate the critical
mutation probability $u_c$ that separates the sub and supercritical regimes. Accordingly, in Figs.\ \ref{fig:5}
and \ref{fig:6} we show log-log plots of $N_t$ and $P_t$ for values of $u$ close to $u_c$. Visual inspection
of Fig.\ \ref{fig:5} leads to the estimate $u_c = 0.185 \pm 0.001$ for case  $d=\Lambda=2$. As pointed out before,
in the subcritical regime one has $N_t \sim \exp \left ( \xi t \right )$ with $\xi \sim - \mid \Delta \mid ^\nu$.
The time decay constant $\xi$ is calculated by replotting Fig.\ \ref{fig:6} in the semi-log scale (see Fig.\ \ref{fig:2}
for a similar graph) and then fitting the  straight lines that result from this scale transformation. Finally, Fig.\ \ref{fig:7}
confirms the linear dependence of $\xi$ on $\mid \Delta \mid$, and so the exponent value $\nu = 1$. We conclude then
that the ultimate survival probability vanishes as $\Pi \sim \Delta$ when the critical
mutation probability is approached from the supercritical regime.
These results assure
us of the reliability and adequacy of the spreading technique to locate the mutation threshold in
the package model of \citet{Niesert1981}. This technical procedure was repeated for different values of 
the control parameters $d$ and $\Lambda$ in order to obtain a complete phase diagram of the model. Finally,
we note that in this scheme we have imposed no limitation whatsoever to the total protocell population size. 


\begin{figure}
\centerline{
\resizebox{0.75\columnwidth}{!}{\includegraphics{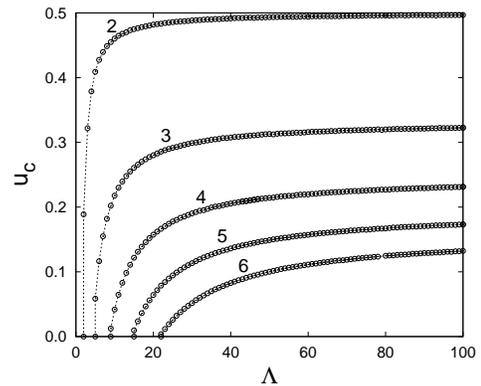}}}
\caption{Critical mutation probability to the parasite class $u_c$ as function of replicase processivity $\Lambda$ 
for values of the template diversity $d$ indicated in the figure. Mutations to the lethal class are no permitted  ($v = 0$).  
}
\label{fig:8}
\end{figure}

Fig.\ \ref{fig:8} summarizes the dependence of the critical mutation probability $u_c$  on $d$ and $\Lambda$.
We recall that for $u \geq u_c$ the lineage is unviable, i.e.,  $\Pi = 0$. For low diversity, the introduction of 
parasites has practically no effect on the viability of the lineage. Solely when the mutation probability
takes very large values then the harm caused by the parasites becomes appreciable
and must be compensated by the increase of the replicase
processivity $\Lambda$. For high diversity, however, the fate of the lineage is much more sensitive to the
presence of parasites, meaning that metapopulations with a high diversity of functional templates must have 
a very efficient replicase right at the beginning, which  happens to be the situation of modern organisms. 
All polymerase cores involved in  replication have similar processivity values, 
regardless of the number of genes per genome, which can vary from few hundreds genes in the smallest 
prokaryotes to tens of thousands in the largest eukaryotic genomes \citep{Benkovic2001}. We note, however,
that the more sundry viral polymerases  do not follow this scenario. 

\begin{figure}
\centerline{
\resizebox{0.75\columnwidth}{!}{\includegraphics{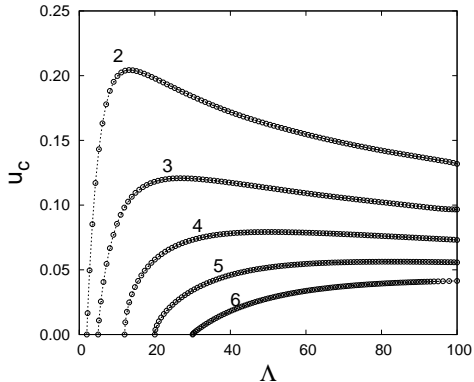}}}
\caption{Same as Fig.\ \ref{fig:8} but using the prospective value of \citet{Niesert1981} to discard
supernumerary protocells. This additional ingredient leads to the incorrect prediction
that high processivity values can be deleterious to the protocell population, in
the sense of curtailing its region of viability. }
\label{fig:9}
\end{figure}

A salient feature of the metapopulation dynamics revealed by Fig.\ \ref{fig:8} is that, for fixed $d$, there is a
value of the mutation probability beyond which coexistence is impossible regardless of the value of the replicase processivity. This
is so because $u_c$ tends to the well-defined value $1/d$ in the limit $\Lambda \to \infty$. Nevertheless, contrary to 
the claim of \citet{Niesert1981}, increasing $\Lambda$ is always benefic to the lineage: there are no `mutational 
reefs' or Charybdis' partner -- the monster Scylla -- awaiting for the metapopulation. In fact, as pointed out before, their conclusion
was a result of the prospective value used to limit the metapopulation size,  which was based on  three properties,
namely, the degree of equipartition  of the copies among the 
different functional templates, the number of parasites and the overall redundancy of the functional templates.
This claim is confirmed by  the result of 
the simulation of the metapopulation dynamics using the prospective value of \citet{Niesert1981} as shown in Fig.\ \ref{fig:9},
from where we see that $u_c$ increases towards a maximum 
and then decreases towards zero as  $\Lambda$  increases further.
Rather surprisingly, the introduction of an explicit group selective pressure towards more balanced protocells turns out
to be harmful for the metapopulation, which now can stand much lower values of the mutation probability.
The reason is that for large $\Lambda$ and not too low $u$, 
the surviving vesicles in the supercritical regime are heavily loaded with parasites and so use of such 
selection criterion would purge them from the population resulting  in the premature extinction of the lineage.

\section{The role of lethal mutants}\label{sec:lethals}

Lethal mutants are precipitating agents, the action of which disrupts the ability of the set of functional templates
to assemble the  replicase. Accordingly and following  \citet{Niesert1981}  we assume that the activity of the lethal mutants
comes about only after the fission of the mother protocell, so that only the daughter protocells are affected
by this type of mutant. This is a best case scenario, since there is a chance that  one of the daughters comes
out free of lethal mutants. 

\begin{figure}
\centerline{
\resizebox{0.75\columnwidth}{!}{\includegraphics{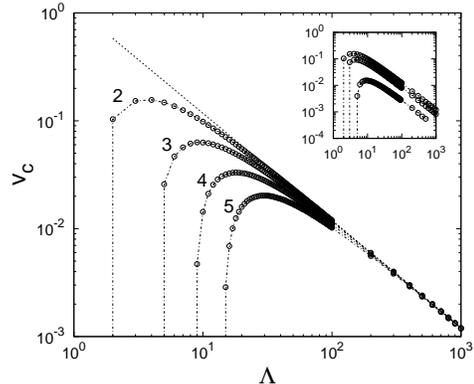}}}
\caption{Critical  mutation probability  to the lethal class as function of the replicase processivity  in the absence 
of parasites ($u=0$)  for the
values of the template diversity $d$ indicated in the figure. For large $\Lambda$ we find $v_c \sim 1/\Lambda $ (dashed
straight line) regardless of the value of $d$. The inset show $v_c$ vs. $\Lambda$ for $d=2$ and (top to bottom) 
$u=0, 0.2$ and $0.4$, indicating that the asymptotic scaling $v_c \sim 1/\Lambda $ is
independent of $u$ as well.}
\label{fig:10}
\end{figure}

\begin{figure}
\centerline{
\resizebox{0.75\columnwidth}{!}{\includegraphics{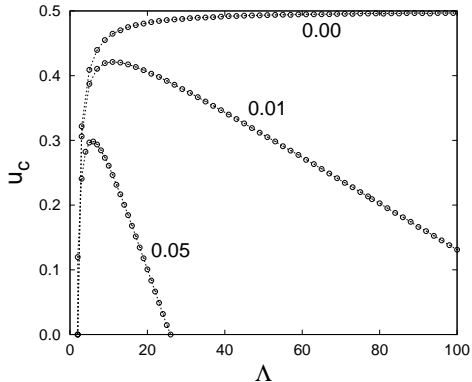}}}
\caption{Critical mutation probability to the parasite class as function of the replicase processivity 
for $d=2$ and $v$ as indicated in the figure. The upper bound on $\Lambda$ is due solely to 
the effect of lethal mutants, whereas the lower limit is due to the assortment load.}
\label{fig:11}
\end{figure}

Since the mutation probability to the lethal class $v$ is a real variable, we can use
the spreading technique to locate the critical value $v_c$ that separates the 
sub and the supercritical regimes.  The main graph of Fig.\ \ref{fig:10} summarizes our results in the case
that mutations to the parasite class are forbidden ($u=0$). In this case it is clear that  too high processivity values
are harmful to the metapopulation. A compromise between the `fluctuation abyss' (loss of functional templates
due to the assortment load) and  the `mutational reefs' (production of lethal mutants) is achieved by a finite
value of $\Lambda$, which corresponds to the maximum in the curves of Fig.\ \ref{fig:10}. It is interesting that
$v_c$ becomes practically independent of $d$ in the limit of large $\Lambda$, where the data is very well fitted
by the function $v_c \sim 1/\Lambda$, which essentially means that the average number of lethal mutants must be smaller than 1
in any  viable lineage. In fact, the relation $v_c \Lambda \approx 1$ holds in the case mutations to the
parasite class are allowed as well, as shown in the inset graph of Fig.\ \ref{fig:10}. The only effect of
$u > 0$ is to shift uniformly  $v_c$ towards lower values. The important point illustrated by
Fig.\ \ref{fig:10}  is that for fixed $v$  the metapopulation is viable only between a 
certain range of processivity values,  which excludes too small as well as too high values of $\Lambda$.
This result is  better illustrated in Fig.\ \ref{fig:11} which shows the dependence of $u_c$ on $\Lambda$ 
for $v$ and $d$  fixed. 

We note that the presence of parasites does not add any qualitative
new feature to the model dynamics:  the threshold phenomenon for low $\Lambda$ is due to the assortment
load only, whereas the threshold for high $\Lambda$ results solely from  the destructive activity of  the lethal
mutants. This is in stark contrast with the major role played by $u$ in free-gene models, as evidenced by 
the error-threshold in the deterministic quasispecies model \citep{Eigen1971} and Muller's ratchet in finite population models  
\citep{Haigh1978}. In fact, the analogue of the error threshold transition 
is the limiting value $u_c \sim 1/d$ for large $\Lambda$ (see Fig. \ref{fig:8}) which
is irrelevant here because the presence of lethal mutants prevents the evolution of replicases with
such high processivity values. Nonetheless, the existence of an error threshold-like transition between 
viable and unviable states with respect to lethality contrasts with the absence of thresholds in the presence of lethal 
mutants in the quasispecies model \citep{Wilke2005} 
and corroborates the findings of \citet{Takeuchi2007}. 
The reason the steadily accumulation of parasite mutants does not lead to
a phenomenon similar to the Muller's ratchet is that  we have assumed a truncated fitness landscape at the protocell 
level, so protocell containing only parasites cannot accumulate. In any event, since the metapopulation
can undergo unrestrained growth in the supercritical regime, the accumulation of parasites is not a
serious hindrance as it is in the case the metapopulation size is fixed (see, e.g., \citet{Fontanari2003}).
\section{Conclusion}\label{sec:conclusion}

The reexamination of the classic package model of \citet{Niesert1981} has revealed 
a few novel features and
clarified some aspects of the metapopulation dynamics, which were obscured in the
original work by the  introduction of an artificial prospective value to limit
the metapopulation size. To circumvent this difficulty, we have resorted to the
powerful spreading technique of statistical mechanics which, by focusing on
the time dependence of a few quantities, permits
the precise location of the parameter values bordering the sub and supercritical regimes
of population growth.  In addition,  our computational resources allow us to handle
metapopulation sizes of order of $10^6$ packages, a figure that would be unthinkable in the
beginning of the 1980s.

In the case of finite processivity $\Lambda$,
which is the situation of interest in package models, we find that
the parasites have no significant role in the evolution of the metapopulation and, in
particular, mutations to the parasite class do not set an upper bond to the values that
$\Lambda$ can assume. This is in disagreement with the findings of the original analysis
(see  Figs.\ \ref{fig:8} and \ref{fig:9}). In fact, the phenomena responsible for the scenario  `life between 
Scylla and  Charybdis' are the assortment load (i.e., the random assignment  of templates to the daughter 
protocells) which sets  a lower bound to $\Lambda$ and the presence of lethal mutants which, in turn, set the
upper bound to $\Lambda$ (see Figs.\ \ref{fig:10} and \ref{fig:11}).

An important result, which seems to have been completely overlooked by \citet{Niesert1981}, regards
the apparently unproblematic situation where the replication fidelity is perfect, i.e., $u=v=0$. The
difficulty is illustrated in Fig.\ \ref{fig:4} which shows  that the lower the diversity, the
higher the growth rate of the metapopulation. So, in the case that two such  
metapopulations are set to compete in the same environment, it is evident that the lineage with higher 
diversity will inevitably be excluded. The situation becomes even worse in the presence of parasites
and lethal mutants, since these mutants tend to cause more harm to the lineages with high diversity (see
Figs.\ \ref{fig:8} and \ref{fig:10}). This is a very unpleasant situation, the solution of which 
requires ad hoc assumptions about the protocell fitness (e.g., a fitness that increases with  the template diversity)
or about the dependence of $\Lambda$ on $d$
in order to revert the negative scenario revealed by Fig.\ \ref{fig:4}.

A word is in order about the amount of extra information   eked out by confining the templates in packages. 
Here we equate information content with template length. The fundamental
quantities here are the spontaneous error rate per nucleotide $\epsilon$, the molecule length $L \gg 1$, and 
the fraction of neutral single nucleotide substitutions $\lambda$ (see, e.g., \citet{Takeuchi2005})
from which we can readily obtain the probability that a template copy becomes nonfunctional 
(i.e., a parasite or a lethal),
\begin{equation}\label{mu}
\mu = u +v =  1 - \exp \left [ - \epsilon \left (1 - \lambda \right) L \right ] .
\end{equation}
A generous estimate
of these parameters (see, e.g., \citet{Drake1999,Johnston2001,Kun2005}) is $\epsilon \sim 0.005$, $L \sim 200$ and
$\lambda \sim 0.25$ which yields $\mu \sim 0.53$. This is a truly disastrous result which precludes the coexistence
of even two templates. Alternatively, we can use Eq.\ (\ref{mu}) to calculate the maximum length $L_m$ of a template in
a best case scenario where lethal mutants are absent $v=0$ and the replicase processivity is infinite so that $u_c \approx 1/d$.
The result is $L_m = - \left [ \ln \left ( 1 - 1/d \right) \right ]/\left [ \epsilon \left ( 1 - \lambda \right ) \right ]$,
which, using  the same parameter values as before, yields $L_m \sim 77$  for $d=4$, resulting thus in a total of about 300 nucleotides.
Considering that this is a best case scenario, the improvement over the free-gene situation (about 200 nucleotides)
is scanty.  For instance,  for large $d$  we have
$L_m d \approx 1/\left [ \epsilon \left ( 1 - \lambda \right ) \right ]$ which is essentially a constant value that
depends on environmental factors only. Hence there is no information gain by increasing $d$ since the length $L_m$ of
the coexisting templates  must decrease in the same proportion so as to keep the total information  content
(i.e., $L d$) constant. A similar `information conservation principle' is likely to hold for the hypercycle
model  as well. We recall that in the case of the elementary hypercycle, the maximum number of templates that can coexist in a
dynamical equilibrium state is $d=4$ \citep{Eigen1978}. The analysis of the error propagation
in the hypercycle \citep{Campos2000} is not useful
here because the error tail (mutant) class considered in that work is not equivalent to the parasite class
of the package models, since those mutants do no receive catalytic support from the hypercycle members.

The original package model proposed by  \citet{Niesert1981} and reviewed  here is truly extraordinary for a 
simple reason, which perhaps passed unnoticed even to their proponents: because the
protocells do not compete and the lineage is free to grow unrestrainedly, the region of viability
of the metapopulation in this model is the greatest possible! Introducing
the notion of protocell fitness or prospective survival value that depends explicitly
on the template composition of the protocells, regardless of 
the prescription one chooses, can only reduce the size of that region. This is the reason
that the critical mutation probability in Fig.\ \ref{fig:8} is greater than its counterpart
in  Fig.\ \ref{fig:9}. This means that our estimates for the critical  mutation probabilities
$u_c$ and $v_c$ are upper bounds to the critical mutation probability of any package model for which 
the survival of the protocell is linked to the coexistence of a set of functional templates. Therefore,
since even in this best case scenario the information gain derived from the coexistence of the
distinct templates is not significant, this class of models should be discarded as possible solutions to
the prebiotic information crisis. This is not to say that packages are not important in prebiotic evolution, 
since it is undisputable that compartments are essential to the  evaluation of the translation products of the
information coded in the templates \citep{Eigen1980}. 

Coming up with a coherent scenario  to explain the coexistence of distinct templates has proved
to be a most difficult endeavor and  it may  already be time to turn to new approaches to solve
the 
information crisis of prebiotic evolution. We have nothing to offer on this direction.

\section{Acknowledgements}
D.A.M.M.S. is supported by CAPES. The work of J.F.F was supported in part by CNPq and FAPESP, Project No. 04/06156-3.

\end{document}